# SENSOR-BASED SPREADER AUTOMATION FOR REDUCING SALT USE AND IMPROVING SAFETY

**Ayushmaan Aggarwal**
Lakeland High School
1349 E. Main Street,
Shrub Oak, NY 10588
Email: ayushmaanaggarwal@icloud.com

**Niharika Bhattacharjee**
Yorktown High School
2727 Crompond Road
Yorktown Heights, NY 10598
Email: monkeymania25@gmail.com

**Aadi Bhattacharya**
Valhalla Middle School
300 Columbus Ave
Valhalla, NY 10595
Email: aadi.gbh@gmail.com

**Raka Bose**
Mildred E Strang Middle School
2701 Crompond Road
Yorktown Heights, NY 10598
Email: polaris1b1@gmail.com

**Anshul Gupta,** *Corresponding Author*
IBM T.J. Watson Research Center
1101 Kitchawan Road
Yorktown Heights, NY 10598
Tel: 914-945-1450; Fax: 914-945-3434; Email: anshul@us.ibm.com

**Deepta Gupta**
Rye Country Day School
3 Cedar Street
Rye, NY 10580
Email: deepta.bg@gmail.com

**Anuj Kapoor**
Lakeland Copper Beech Middle School
3417 Old Yorktown Rd,
Yorktown Heights, NY 10598
Email: anuj19kapoor@gmail.com

**Elina Rani**
Lakeland High School
1349 E. Main Street,
Shrub Oak, NY 10588
Email: elinaranicty@gmail.com

**ABSTRACT**

Over 30 million tons of deicing salt is applied on U.S. roads annually at a cost of roughly $1.2 billion and with significant negative environmental impact. Therefore, it is desirable to reduce salt use while maintaining winter road safety. Automatic adjustment of application rate in response to road, weather, traffic, and other conditions has the potential to achieve this goal.

In the US, salt application rates are typically pre-set manually based on roadway classification, cycle time, desired level of service (LOS), and expected traffic, road, and weather conditions. The operators can temporarily change the application rate manually based on their experience and observations. Current spreader automation mostly involves adjusting discharge rate in response to spreader speed, although pavement temperature sensors are likely to be adopted in the future.

This paper explores extending spreader automation for adjusting salt discharge rate on curves, inclines, and other areas with historically high accident concentrations. First, we propose the use of gyroscopic sensors and inclinometers on board the spreader to adjust discharge rate in real time in response to curves and inclines. The second part of the proposed solution includes the installation of roadside Radio-Frequency Identification (RFID) tags that can communicate the length and the severity of the hazard zone to an RFID reader on board the spreader to automatically adjust discharge rate accordingly.

Our proposed methods have the potential to reduce salt use and operator fatigue, while increasing winter road safety without adding excessive cost or complexity to the existing spreader systems in the US.

# 1. INTRODUCTION

Over 30 million tons of deicing salt is applied on the US roads every winter at a cost of roughly $1.2 billion and with significant negative environmental impact [1,2,3]. As a result, transportation agencies, such as state departments of transportation (DOTs), are trying to reduce their salt use without compromising driver safety [4].

Automatic adjustment of application rate in response to road, weather, traffic, and other conditions has the potential to reduce salt use while maintaining or enhancing safety for the vehicles on the road [5] and for the snow plow equipment operators [6]. In the US, application rates are typically pre-set manually based on roadway classification, expected traffic, road, and weather conditions, cycle time, and the desired level of service (LOS). Often, the operators can temporarily increase the application rate manually based on their experience and their observation of the road conditions. Current spreader automation in the US mostly involves adjusting discharge rate based on spreader speed, although pavement temperature sensors are considered highly desirable and are likely to be adopted in the future [5].

In this paper, we explore relatively simple and inexpensive spreader automation to prevent or minimize overapplication of salt due to manual rate adjustment during the beat. In our proposed method, the spreader would increase discharge rate automatically on portions of a road where accidents are more likely to occur, such as curves, inclines, and other areas with a history of a higher concentration of accidents. We propose using gyroscopic sensors and inclinometers on board the spreader to detect the sharpness of curves and the steepness of inclines in real time and adjust the discharge rate automatically in response [7]. We also propose installation of stationary roadside Radio-Frequency Identification (RFID) tags near known accident-prone zones. These tags can convey the length and the severity of the hazard zone to an RFID reader on board the spreader to automatically adjust discharge rate and increase the LOS in such zones.

The proposed gyroscopic sensors, inclinometers, and RFID readers are inexpensive and can be connected to existing spreader controllers via the CAN (Controller Area Network) bus, which is standard on the spreaders. Enhancing the controllers' software to take the additional sensor and RFID input into account would complete the proposed automation. Our proposed methods have the potential to reduce salt use and operator fatigue, while increasing winter road safety without adding excessive cost or complexity to the existing spreader systems in the US.

In this paper, we also make the case that an RFID-based solution [8] might be able to reduce some of the cost and complexity and improve accuracy in some of the advanced GPS-based automation that is currently being pursued in Europe [9,10]. Transportation agencies in the US may prefer sensor- and RFID-based automation in the future, because of its simplicity and lower cost.

## 2. PROBLEM

Optimum salt application rate depends on many factors, including pavement temperature and road geometrics. According to snow and ice control guidelines [11] from New York State Department of Transportation (NYSDOT), the factors that influence application rates include steep grades, sharp curves, bridge decks, and cold spots, etc. Some of these situations determine the application rate for a whole beat, and others require the drivers to make adjustments during their run. The reason is that the goal of normal spreading operations is not to achieve black pavement, but to achieve effective anti-icing; i.e., preventing the snow or ice from adhering to the pavement. However, even a small amount of loose snow or ice could be hazardous on curves and inclines and the operators may decide to increase the LOS on such portions of the roadway.

Increasing the application rate or the LOS for an entire beat to compensate for hazardous portions on the route likely results in over-application, because even the safer straight and flat portions of the route are subject to the increased application rate, beyond what is necessary for a successful anti-icing operation. Manual adjustments are also likely to result in over-application, because the operators are likely to err on the side of safety. The current practice of using the "blast" button increases the discharge rate dramatically for a short period of time (usually around 10 seconds) and may result in excess salt application in an area where a smaller increase would have sufficed. In addition, manual adjustments for expected pavement temperature based solely on operator experience can be tricky, as the location and intensity of cold spots depend on a number of factors. Finally, frequent application rate adjustments on a beat can result in distraction and fatigue for the operator, which appears to be a matter of concern [12].

## 3. POSSIBLE SOLUTIONS

We believe that spreader automation can reduce the overall amount of salt used, while maintaining or improving safety. Automation is also likely to reduce operator fatigue and improve the safety of snow removal or deicing operations by letting spreader drivers focus on vehicle operation, instead of manually adjusting the application rate [6]. A lot of manual adjustments to the application rate on a beat are made by the operators based on their experience with the route. Spreader automation can help make snow removal or deicing operations more flexible, because with automation, any driver would be able to handle any route. In this section, we propose a number of simple and inexpensive sensor-based automation techniques. Almost all sensors, including RFID readers, can communicate with a spreader's controller over the CAN bus either directly, or through an inexpensive intermediate microprocessor system.

### 3.1. Gyroscopic Sensors, Inclinometers, and Temperature Sensors

Current automation in most spreaders in the US takes only the speed of the spreader truck into account. Operators set a default application rate mostly based on their experience with the route and on expected road, weather, and traffic conditions [11]. If the operator sets the desired application rate at *A* pounds/mile and the spreader is moving at a speed of *s* miles/hour, then salt is discharged at the rate of *d* pounds/hour based on the simple equation:

$$d = s \cdot A. \qquad (1)$$

In the near future, it is expected that automation in US spreaders will include adjusting the discharge rate based on pavement temperature as well. Inexpensive and accurate infrared pavement temperature sensors are readily available. If $\Delta T$ is the difference between the expected and actual pavement temperature, then the original equation (1) can be modified as follows:

$$d = s \cdot A \cdot (1 + k_1 \Delta T). \qquad (2)$$

If the pavement is colder than expected, then $\Delta T$ would be positive, causing the discharge rate to increase. If the pavement temperature is higher than expected, then $\Delta T$ would be negative, causing the discharge rate to decrease. The constant $k_1$ determines how sensitive discharge rate *d* is to $\Delta T$.

In addition to pavement temperature, we also propose detecting and measuring the curves and grade of the roadway in real time. Inclinometers on the spreader truck can measure the steepness of slopes, and gyroscopic sensors can detect sharpness of curves by measuring the truck's angular velocity. If $\theta$ is the angle of incline or decline of the roadway, and $\omega$ is the spreader's angular velocity, then the spreader's controller can calculate discharge rate using the following equation:

$$d = (s + k_3|\omega|) \cdot A \cdot (1 + k_1 \Delta T) \cdot (1 + k_2|\theta|). \qquad (3)$$

The constants $k_2$ and $k_3$ determine the relative weights of grade and sharpness of a curve in increasing the discharge rate. Note that in equation (3), we consider linear dependencies between discharge rate and $\Delta T$, $\theta$, and $\omega$. We use absolute values of $\theta$ and $\omega$, because we consider only the magnitudes of the grade and curve. It is also possible to come up with different equations with nonlinear dependencies and different weights for inclines and declines. The equation must revert to the original $d = s \cdot A$ when $\Delta T$, $\theta$, and $\omega$ are 0.

### 3.2. RFID Devices

Some areas of high accident concentrations have hazards that cannot be detected easily using sensors. Examples of these include highway sections near high-volume exits and entrances, roadway sections with narrow or no shoulder, sections with seasonal morning or evening solar glare, etc. To enhance automation in salt spreaders to take such situations into account, we propose the use of inexpensive and proven RFID technology [8]. Our solution includes having the spreaders equipped with RFID readers and installation of compatible roadside RFID tags at the beginning of known accident-prone sections without readily detectable hazards such as curves, inclines, or cold spots. A roadside RFID tag would transmit simple instructions to a passing spreader, such as numerical values indicating the severity and the length of the accident-prone section. These instructions would direct the spreader to alter the effective application rate by a situation-specific percentage over a specific distance or time. Similarly, RFID tags can direct spreaders to reduce application rates in environmentally sensitive sections, or stop the application on bridges that are treated using alternative chemicals or methods.

## 4. EXPERIMENTAL EVALUATION

We experimentally evaluated part of our solution on a 10-mile stretch of Taconic State Parkway (TSP) from Commerce Street in Valhalla, NY to Route 134 in Ossining, NY. We describe our experiments and insights in the following subsections.

### 4.1. Parameter Setting

We used reasonable guesstimates for the constants $k_1$, $k_2$, and $k_3$ for the discharge rate equation (3) in Section 3.1. Based on a simple linear regression on application rates suggested for different temperatures in various guidelines, including those by NYSDOT [11] (Appendix C), we arrived at a value of 0.05 for $k_1$. This means, that for every 1°F change in temperature, the discharge rate increases by about 5%. However, the focus of our experiments was road geometrics, so we assumed $\Delta T$ to be 0, and the value of $k_1$ did not factor into the calculations.

We initially treated inclines and declines differently, with a higher weight for declines. We used $k_2 = 0.04$ for inclines (when $\theta$ was positive) and $k_2 = 0.08$ for decline (when $\theta$ was negative). This translated to roughly 27.5% increase in discharge rate on a 6% downhill grade, which is equivalent to a $\theta$ of -arctan(0.06) or -3.43°. The value of $k_3$ was set to 2.5 for increasing the discharge rate based on the angular velocity $\omega$. We discuss the relationship between speed $s$, angular velocity $\omega$, and roadway curvature in more detail in Section 5.1.

### 4.2. Initial Testing

We built a simple setup consisting of gyroscopic sensors and a microcontroller to simulate a spreader going over the highway. We measured and recorded real-time incline and angular velocity data. Using this data and the final discharge rate equation (3) in Section 3.1, we calculated the effective application rate every two feet. Figure 1 contains a visual representation of the incline, angular velocity, and the calculated effective application rate on this stretch of highway. The third bar from the left shows how the effective application rate increases from its baseline value in response to changes in incline and angular velocity, which are depicted by the first two bars.

NYSDOT guidelines [11] suggest a general application rate of 160 pounds/lane-mile of pre-wetted salt under light or moderate snow with temperatures above 23°F. Since the test portion of Taconic State Parkway (TSP) has curves and grades, and has a relatively high accident rate, an operator may find it safer to apply salt at 200 pounds/lane-mile, resulting in 2000 pounds of salt use per lane over the 10 miles. Assuming a fixed spreader speed and $\Delta T = 0$, our system was able to vary the application rate between 150 and 350 pounds/lane-mile, and yet use only about 1700 pounds of salt per lane over the 10-mile stretch. The extreme rate of 350 pounds/lane-mile was used only on the ramps, mimicking the use of the manual "blast" button.

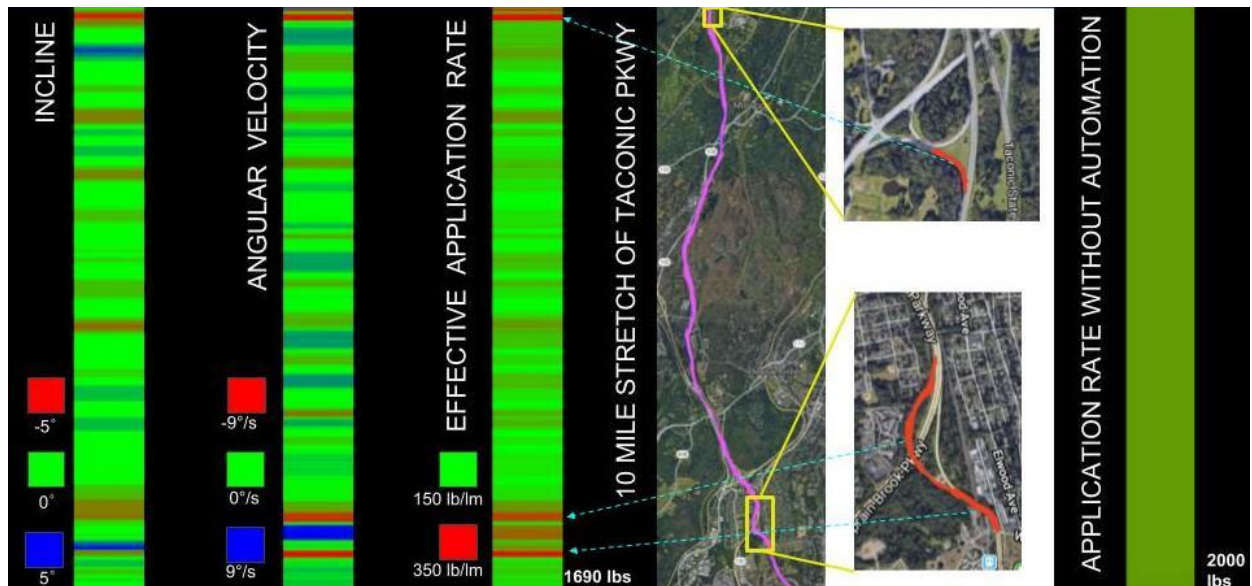

Figure 1: Real-time incline and angular velocity measurements and effective application rate calculations based on our solution over a 10-mile northbound stretch of Taconic State Parkway (TSP), NY.

### 4.3. Preliminary Accident Analysis

We obtained 20 years of accident location data from NYSDOT for our test stretch. From this data, we extracted the locations of all the accidents that occurred when the pavement had slush or snow and mapped these locations alongside our incline and angular velocity measurements and effective application rate calculations. For better clarity, we divided our test stretch into three segments from north to south and in order of increasing overall traffic volume. These segments are shown in Figures 2(a), 2(b), and 2(c). Each red dot on the map represents a single accident under slushy or snowy conditions between 1998 and 2017.

Although more data and more rigorous statistical analysis would be required to calculate the exact correlation of winter-weather accidents with curves or inclines, a visual inspection of our accident maps suggests a strong link. Many areas of relatively higher concentration of accidents are located where there is a sharp curve, a steep grade, or both. As a result, these are also the areas, where our calculated effective application rate is higher. Many of these areas are marked by dotted lines in Figures 2(a), 2(b), and 2(c).

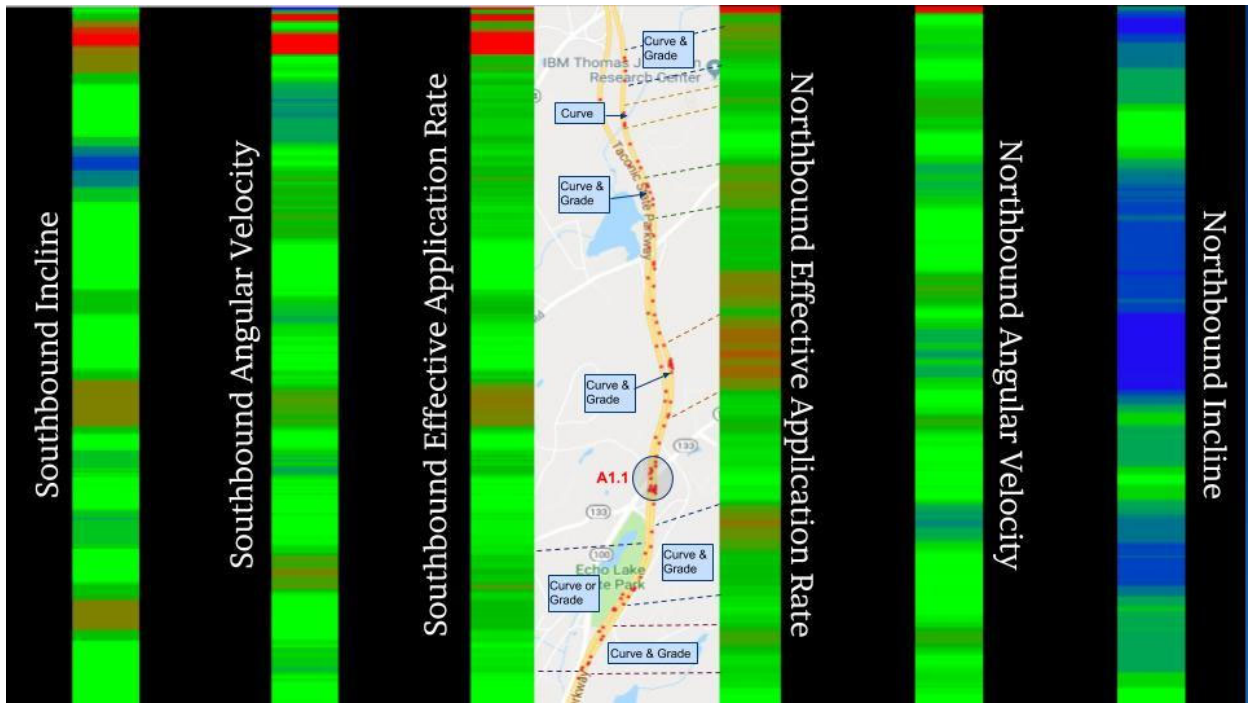

Figure 2(a): Accident map with our initial calculated effective application rate on the northern one-third of the TSP test stretch.

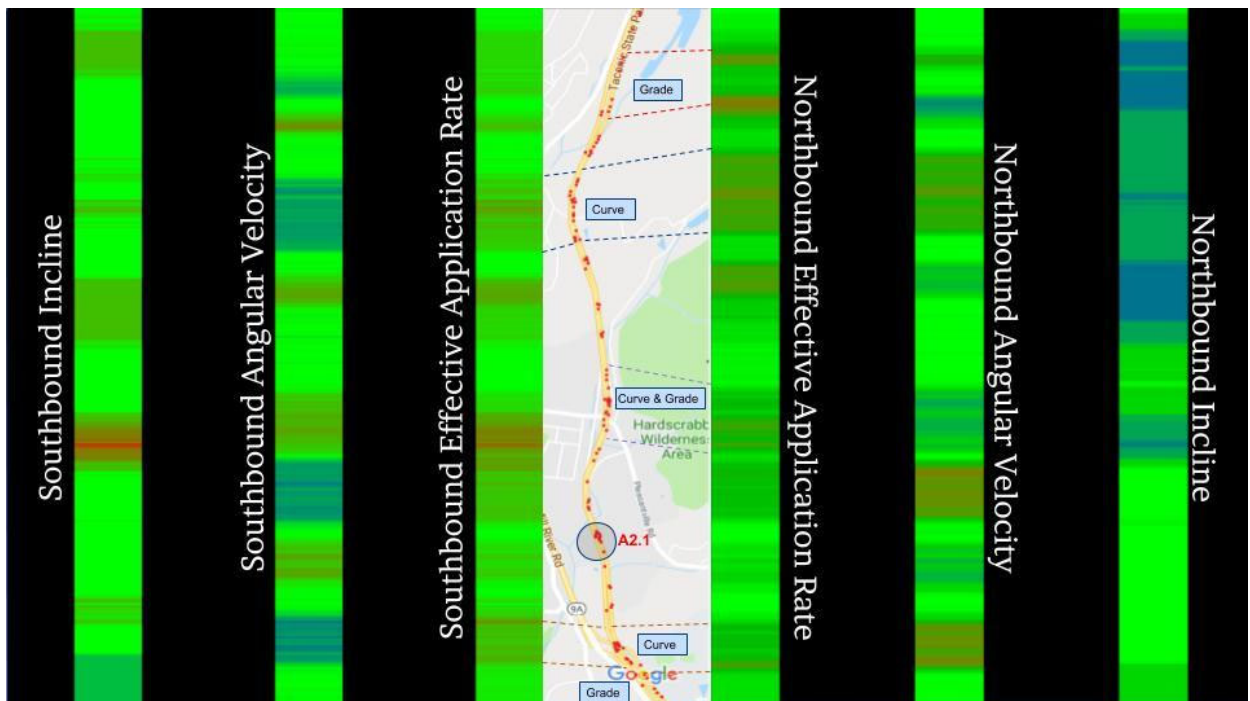

Figure 2(b): Accident map with our initial calculated effective application rate on the middle one-third of the TSP test stretch.

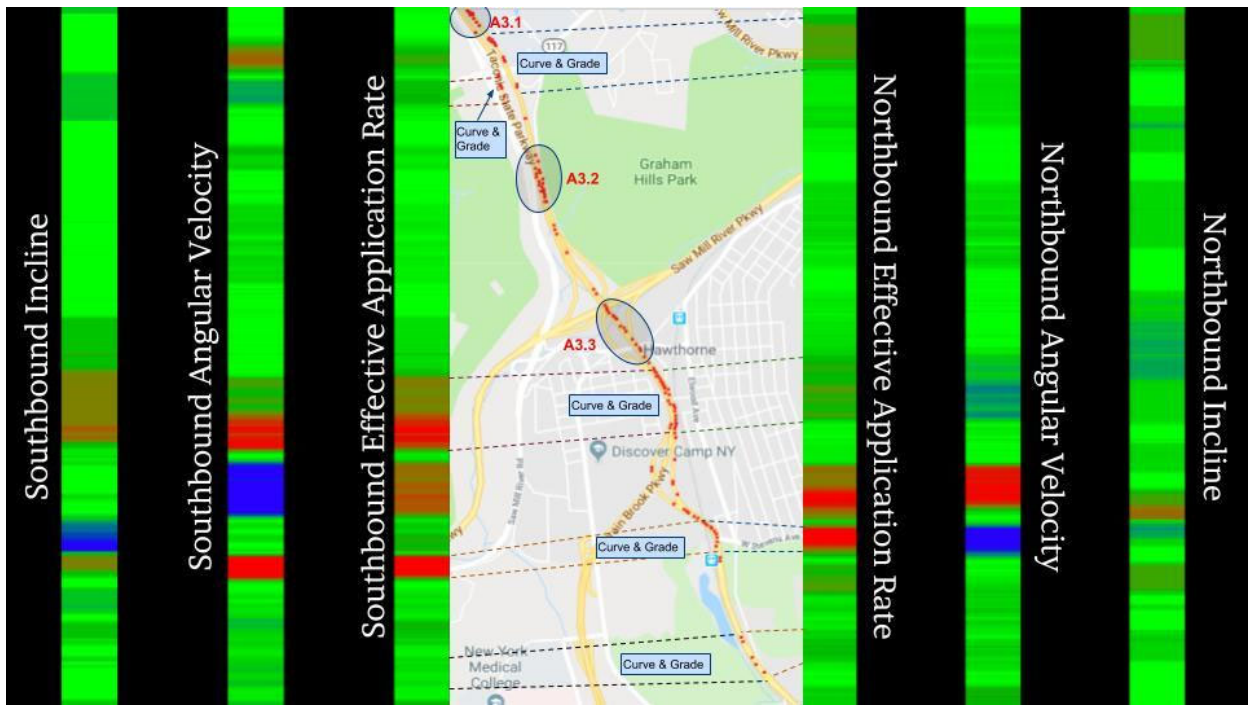

Figure 2(c): Accident map with our initial calculated effective application rate on the southern one-third of the TSP test stretch.

In our accident maps, we also observed some areas of high accident concentration on relatively straight and flat portions of the highway. In our figures, these areas are shown with slightly shaded ellipses. These areas were the motivation behind our RFID-based automation idea described in Section 3.2, which can compensate for the inadequacy of sensor data in such areas.

Closer observation shows that some of these areas are near high-volume exits and entrances, where slowing or merging traffic and lane changes could be contributing factors. Examples of these include A3.1 (exit for Route 9) and A3.2 (exit for Route 117) in Figure 2(c). Some of these accident-prone segments are on bridge decks, where cold spots or a lack of adequate shoulder could be contributing factors. Examples of these include A1.1 (Route 133 overpass) in Figure 2(a) and A3.3 (Sawmill Parkway exit and overpass; no shoulder) in Figure 2(c). One area, A2.1 in Figure 2(b), does not seem to have any obvious hazard.

### 4.4. Improvements Based on Accident Data

After mapping the accidents and comparing their concentrations with our calculated effective application rates, we changed the value of $k_2$ (from 0.04 for incline and 0.08 for decline) to 0.06 for both inclines and declines. With this modification, we improved the correlation between the effective application rate and the concentrations of accidents that occurred under wintry road conditions. Thus, studying the accidents helped us improve our discharge rate model. Figures 3(a), 3(b), and 3(c) show the effective application rates on our test stretch of TSP with the modified constants. A comparison of the northbound sides in Figures 2(a) and 3(a) shows that increasing the weight for incline had the effect of discharging more salt in the most accident-prone zones in the section of highway covered in these figures.

Figures 3(a), 3(b), and 3(c) show that accident data can be used to tune the constants that decide the weightage of incline, decline, and angular velocity. This can allow for additional safety to be taken into consideration because more salt will be spread in the areas of higher accident concentration. We also noticed that our change did not increase the total amount of salt discharged; it only improved the distribution of salt.

### 4.5. Additional Insights from Accident Data

We tuned the weights based on only a visual inspection of accident maps. More data and more rigorous statistical techniques can be used for deriving a more precise equation for calculating effective application rate as a function of curves, inclines, and declines. We also noted that on our test stretch, there appear to be more accidents on the northbound side on an average than on the southbound side. This could be related to the timing of rush hours, which are at or after dusk on the northbound side, but well after dawn on the southbound side during a large part of the snow season. Also, drivers are likely to be more tired during the evening commute than in the morning.

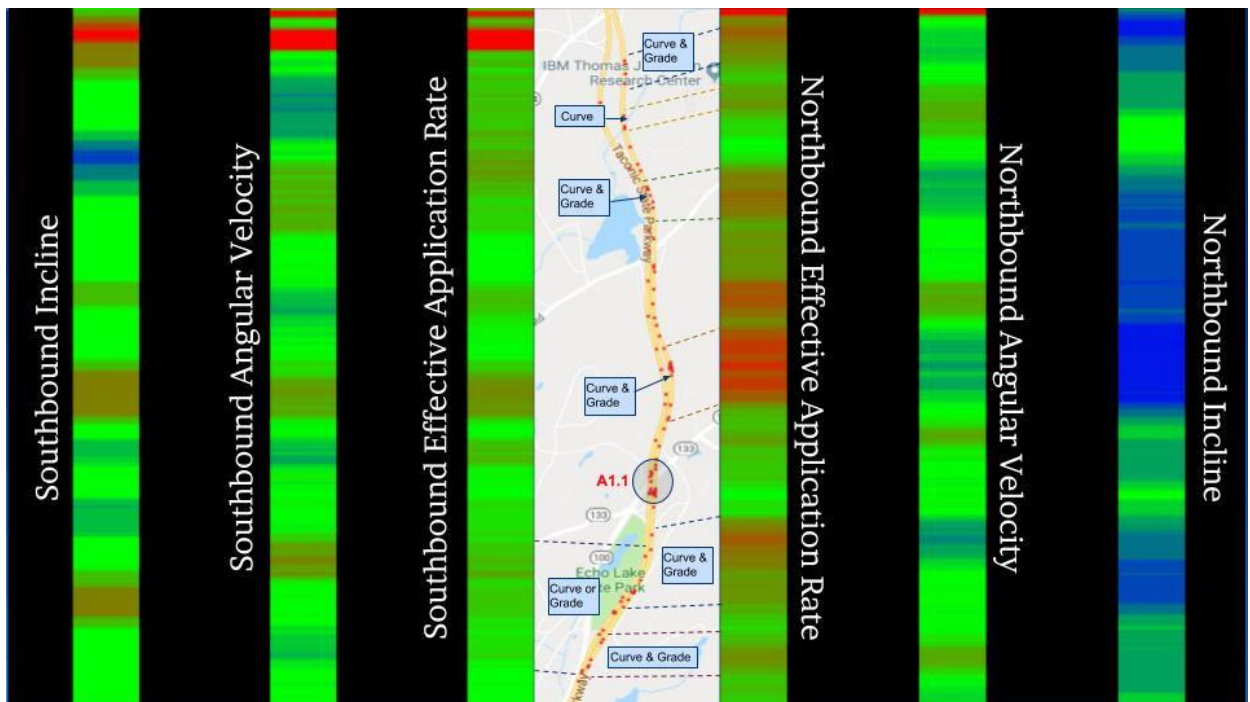

Figure 3(a): Accident map with our modified calculated effective application rate on the northern one-third of the TSP test stretch.

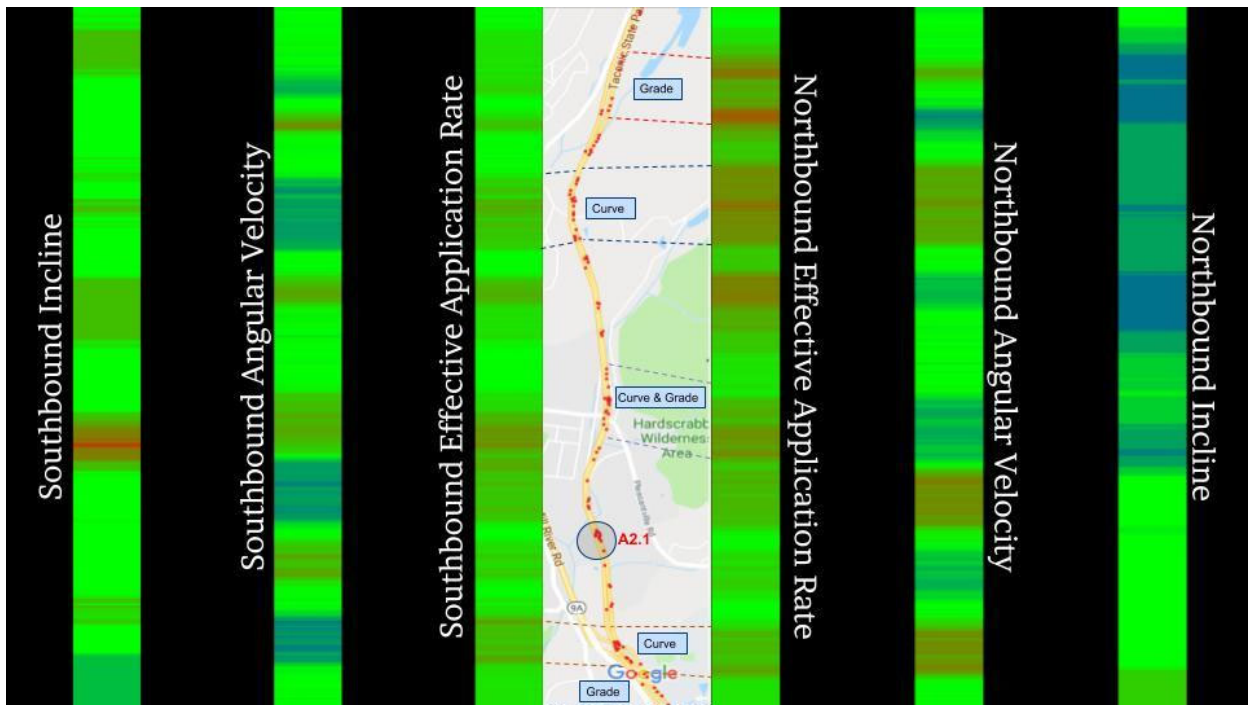

Figure 3(b): Accident map with our modified calculated effective application rate on the middle one-third of the TSP test stretch.

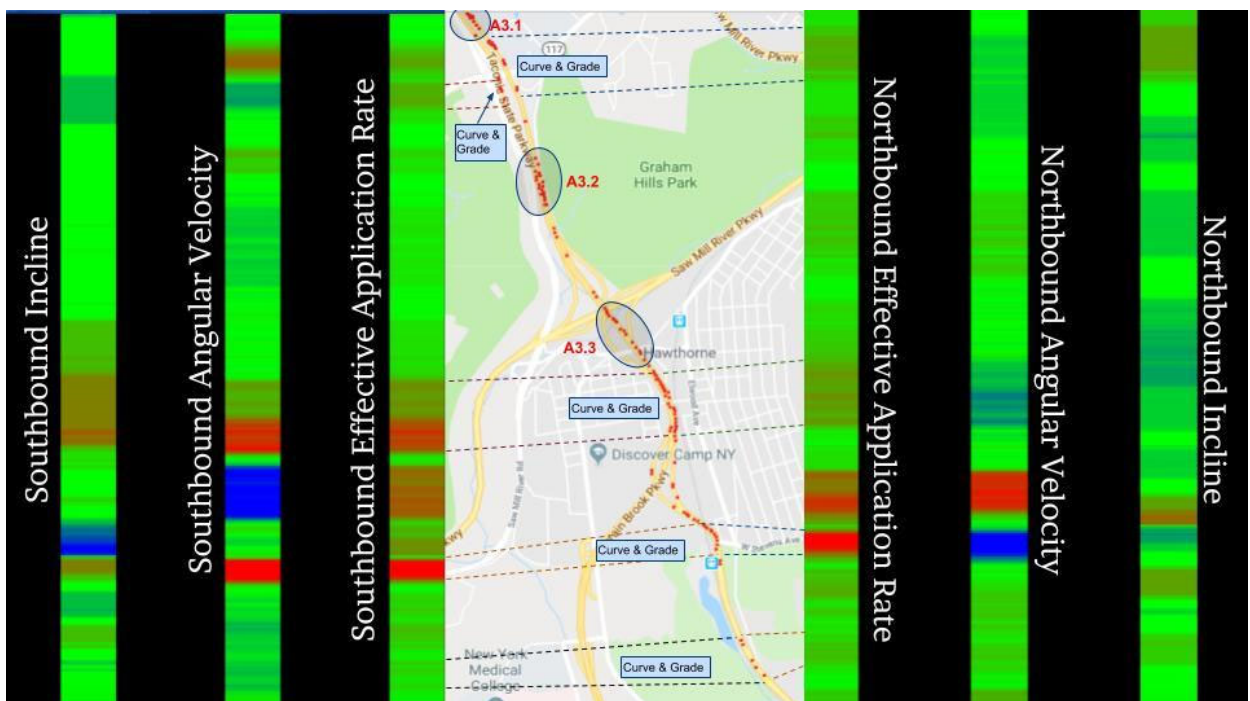

Figure 3(c): Accident map with our modified calculated effective application rate on the southern one-third of the TSP test stretch.

### 4.6. Noise Reduction in Sensor Data

The incline and angular velocity measurements from inclinometers and gyroscopic sensors may be susceptible to errors due to road bumps or spreader vibrations. We did not observe this problem in our experiments because the sensors were attached to a car and were cushioned. On board a spreader, the sensors may require better cushioning and the application of some noise reduction technique to the data they generate. Since typical gyroscopic sensors and inclinometers can generate one reading every 10 milliseconds, we believe that using moving averages would eliminate most noise without affecting precision. If needed, more advanced noise reduction techniques, such as Kalman filter algorithms, can be used.

## 5. ADDITIONAL APPLICATIONS

Some of the proposed solutions presented in this paper have applications beyond real-time adjustment of deicer application rate based on pavement temperature and road geometrics.

### 5.1. Sensors for Fine-Grain Incline and Curvature Data

Inclinometers and gyroscopic sensors on board a vehicle can be used for collecting and recording fine-grain road geometrics data. This data can be used to study the impact of geometrics on accidents and for deriving application rate adjustment formulae based on sensor readings. There are likely to be applications of this data beyond snow removal/deicing operations as well.

The inclinometer can yield incline data every few feet, although the granularity can be increased if desired. The angular velocity measured by the gyroscopic sensors can be combined with the vehicle's real-time speed to obtain lane curvature. Curvature $\kappa$, a measure of the sharpness of a curve, is simply the ratio $\omega/s$, where $\omega$ is the angular velocity and $s$ is the speed of the vehicle in appropriate units. Statistical analysis to relate the incline and curvature data with accident data can be used to determine the role of road geometrics in accidents under various pavement conditions. This information can be used to formulate quantitative guidelines for adjusting deicer applications rates for future automated spreaders.

### 5.2. RFID for Spreader Automation

RFID technology can be useful in spreader automation beyond automatic application rate adjustment in known accident-prone areas. These use cases include the ability to handle temporary or nonstationary hazards and the ability to automate changes in spreading width, pattern, or material.

RFID tags can be placed to change the effective application rate over a portion of a road in response to temporary situations. For example, the glare of the sun can reduce forward visibility, which can be particularly dangerous if braking distance is compromised by slippery pavement. RFID tags can be placed at the beginning of areas with glare only during the days when glare is a problem. The tags can be moved with the location of glare as the angle of sun changes with the calendar. Other example of a temporary hazard is shoulder closure at a work area.

RFID technology can also be used to change the width of spreading, if the road's width or number of lanes change. This is illustrated in Figures 5 and 6. In these figures, the beige color represents the road, where salt is spread and the blue color represents an area, where salt isn't applied. In Figure 5, the image on the left shows, what would happen when a spreader truck drives without changing the spreading width. Some European spreaders [9,10] use sophisticated GPS-based recorded route technology to address such situations. The use of such technology would result in the more desirable spreading pattern shown in the image on the right in Figure 5.

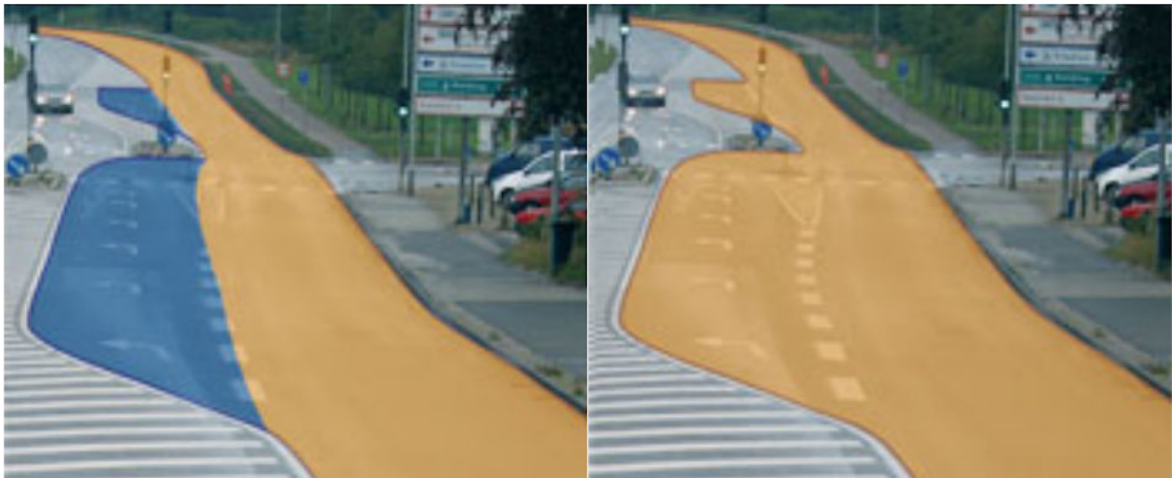

Figure 5:[1] Left: No automation; Right: GPS-controlled recorded route automation for width adjustment.

GPS has an accuracy of about 3 meters, and its use requires customized GIS data specific to the spreader's route. Some of the cost, complexity, and accuracy issues associated with GPS-based recorded route automation can potentially be eliminated by using RFID devices, as shown in Figure 6. Roadside RFID devices can communicate with the RFID reader on the spreader and instruct it to change the width or pattern of spreading for specific distances.

Similarly, RFID technology can be used to instruct a spreader to change the material being spread as needed, for example, on bridge decks. The use of RFID does not require GIS data or pre-recording a route.

---

[1] Picture courtesy [5] with permission.

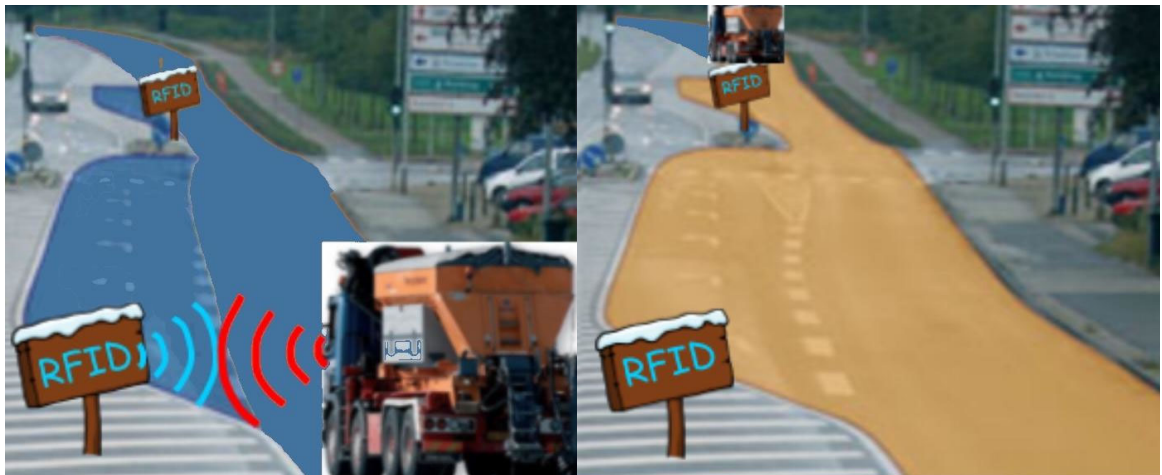

Figure 6:[2] Novel RFID-based width-controlled spreading. Blue area is without salt; beige area has had the salt applied.

## 6. RECOMMENDATIONS FOR FUTURE WORK

Based on our work and preliminary results, we would like to suggest some possible directions of future research for automating and fine-tuning application rates to reduce salt use and improve safety. Our recommendations can be grouped into two broad categories.

### 6.1. Accounting for Road Geometrics

Although some transportation agencies such as NYSDOT suggest that road geometrics have an effect on application rates [11], we did not find any quantitative guidelines for adjusting application rates based on road geometrics. We performed a limited analysis that shows that further research on this topic may lead to improvements in both manual and automatic application rate settings and adjustments.

Our work shows that inclinometers and gyroscopic sensors can be used for improving the efficiency of salt spreading operations. First, such sensors on board a vehicle can be used for collecting fine-grain road geometrics data to study the impact of geometrics on accidents and for deriving application rate adjustment formulae based on sensor readings. Secondly, inclinometers and gyroscopic sensors on board spreader trucks can be used to adjust the application rate in real time based on these formulae. We recommend further research in both these directions.

### 6.2. Use of RFID Technology

As discussed in Section 3.2 and Section 5, RFID technology can play an important role in improving the efficiency of salt spreading operations. Either passive or active (with solar-charged batteries) RFID tags can be inexpensively installed permanently around areas that need adjustment of application rate, width, pattern, or material. If needed, these tags can be easily reprogrammed to adjust to a change in conditions at a given location. In addition, RFID technology is attractive because temporary or mobile tags can help increase LOS precisely in response to temporary situations such solar glare (whose presence and location is highly calendar dependent) or temporary constructions or lane closures. Clear Roads recently conducted a detailed study of spreader automation that included a variety of methods and multiple levels of automation [5], and a cost benefit analysis of various winter road maintenance practices [13]. We recommend that the use of RFID technology be considered in future research on automation and cost-benefit analysis.

---

[2] Picture adapted from [5] with permission.

## 7. CONCLUSIONS

In this paper, we present several proposed solutions that we hope are of interest to transportation agencies such as NYSDOT, spreader manufacturers such as Henderson, Swenson, Aebi-Schmidt, and Epoke, etc., and research consortiums like Clear Roads.

In Sections 3, we show how inclinometers and gyroscopic sensors can be used to improve spreader automation. In Section 4, we demonstrate that in conjunction with historical accident location data, these sensors can help refine application rate guidelines to quantitatively incorporate road geometrics. We hope that DOTs will investigate and conduct future research in this area. Further, we have shown how this can be done without using a complex setup.

Spreader automation in Europe appears to be more advanced than in the US [9,10]. We believe that some of this advanced automation can be achieved in the US with lower cost and complexity using onboard sensors and RFID technology. We hope that spreader manufacturers, transportation agencies, and research organizations would consider looking into the feasibility of using RFID technology in winter road maintenance operations.

## 8. ACKNOWLEDGEMENTS

We would like to thank Mr. Michael Lashmet, Mr. Joe Thompson, and Mr. Edward Goff of NYSDOT for detailed information about winter road maintenance in NY State. Mr. Andrew Sattinger of NYSDOT helped us obtain 20 years of accident location records for our test portion of Taconic State Parkway. We are thankful to Mr. Shane Chesmore, Vice President of Engineering at Henderson Products, USA for technical information about the various aspects of their spreaders. We would like to acknowledge the help from Mr. Arjan Ruiterkamp and Ms. Manon Proksch-Wanders of Aebi-Schmidt for technical information about some of the spreader automation in Europe. We would like to thank the Thompson Engineering Company for allowing the team to use images from their reports. Many thanks to Dr. Michael Dietz of University of Connecticut, and Dr. Stuart Findlay and Ms. Vicky Kelly of the Cary Institute of Ecosystem Studies for feedback on our research and presentations.TRB 2019 Annual Meeting                                                                                      Paper revised from original submittal.

## 9. AUTHOR CONTRIBUTION

The authors confirm contribution to the paper as follows: *Study conception and design:* Ayushmaan Aggarwal, Niharika Bhattacharjee, Aadi Bhattacharya, Deepta Gupta, Elina Rani; *Data collection:* Ayushmaan Aggarwal, Deepta Gupta, Aadi Bhattacharya, Raka Bose, Deepta Gupta, Anuj Kapoor, Elina Rani; *Analysis and interpretation of results:* Ayushmaan Aggarwal, Niharika Bhattacharjee, Deepta Gupta, Elina Rani; *Draft manuscript preparation:* Ayushmaan Aggarwal, Niharika Bhattacharjee, Aadi Bhattacharya, Raka Bose, Anshul Gupta, Deepta Gupta, Elina Rani. All authors reviewed the results and approved the final version of the manuscript.

## REFERENCES

1. Michael E. Dietz and Lukas A. McNaboe. *Road Salt Use in Connecticut: Understanding the Consequences of the Quest for Dry Pavement.* University of Connecticut, February 2017.
2. Victoria R. Kelly, et al. *Road Salt: Moving Towards the Solution*. Cary Institute of Ecosystem Studies, December 2010.
3. *Road Salt and Water Quality.* New Hampshire Department of Transportation, August 2016.
4. Michael H. Lashmet II. *Snow and Ice Control in New York State*. Lake George Park Commision Forum, New York State Department of Transportation, April 2013.
5. *Developing a Totally Automated Spreader System*. Project 99392, Clear Roads (clearroads.org), February 2014.
6. *Automated Spreading Systems for Winter Maintenance: Research Brief*. Clear Roads (clearroads.org), August 2014.
7. Ayushmaan Aggarwal, et al. *Method and system for controlling application rate of de-icing and anti-icing products on the road surface based on sensors on-board the spreading vehicle.* Patent pending. USPTO application no. 62/632,413. February 19, 2018.
8. Ayushmaan Aggarwal, et al. *Method, system, and computer program for altering the discharge rate, discharge pattern, and material of road surface applicants based on temporary or permanent conditions.* Patent pending. USPTO application no. 62/645,140. March 19, 2018.
9. *Autologic Spreading System*, ASH Group, https://www.aebi-schmidt.com/en/products/289/295.
10. *EpoSat GPS-Controlled Spreading*, Epoke, http://www.epoke.dk/home/products/technology/eposat/.
11. *Snow and Ice Control: Highway Maintenance Guidelines Chapter 5*. Operations Division, Office of Transportation Maintenance, New York State Department of Transportation, April 2006 (Revised January 2012).
12. *Identification and Recommendations for Correction of Equipment Factors Causing Fatigue in Snowplow Operators*. Project 1001325, Clear Roads (clearroads.org), June 2016.
13. David Veneziano, Laura Fay, and Xianming Shi. *Development of a Toolkit for Cost-Benefit Analysis of Specific Winter Maintenance Practices, Equipment and Operations*. Clear Roads (clearroads.org), August 2013.

## ABOUT THE AUTHORS

Ayushmaan Aggarwal (grade 10), Niharika Bhattacharjee (grade 10), Aadi Bhattacharya (grade 8), Raka Bose (grade 8), Deepta Gupta (grade 10), Anuj Kapoor (grade 7), and Elina Rani (grade 10) belong to a *FIRST*[R] Lego[R] League (FLL) robotics and STEM team called the Hotshot Hotwires. This paper is the result of the Hotshot Hotwires' research project for the 2017-18 FLL season. Their project, aimed at mitigating the growing problem of contamination of fresh water sources by road salt, was nominated for the FLL Global Innovation award by Hudson Valley FLL. The Hotshot Hotwires were one of the five teams to represent NY State at the FLL World Festival in Detroit, MI in April 2018, where they were among the 13 teams (five from the US) to win top awards.